\documentclass[12pt]{iopart} 
\usepackage{float}
\usepackage{iopams}  
\usepackage{graphicx}
\begin{document}
\title{The influence of velocity field on simple chemical reactions in viscous flow}
\author{A.R.~Karimov$^{1,2^*}$, M.A. Taleisnik$^{3}$,  T.V. Savenkova$^{3}$, L.M.~Aksenova$^{4}$}
\address{$^1$Institute for High Temperatures,
Russian Academy of Sciences, Izhorskaya 13/19, Moscow 127412, Russia\\
$^2$National Research Nuclear University MEPhI, Kashirskoye shosse 31, Moscow, 115409, Russia \\
$^3$All-Russian Scientific Research Institute of Confectionery Industry - Branch of V.M.~Gorbatov Federal Research Center for Food Systems RAS 
Electrozavodskaya 20, Moscow 107023, Russia\\
$^4$V.M. Gorbatov Federal Research Center for Food Systems, Russian Academy of Sciences, Talalikhin 26,   Moscow 109316, Russia\\
$^*$email: arkarimov@mephi.ru}
\begin{abstract}
The nonlinear dynamics of viscous, flow in a compressible spatially homogeneous fluid with chemical reactions are studied. Time-dependent but exact solution of the Piseuille type is obtained. Proceeding from this solution, the influence of a viscosity effect and a chemical reaction on the characteristics of nonequilibrium temporal states of the system is considered. Special attention is drawn to the effect of density collapses on the development of chemical reactions. 

\end{abstract}

\section{Introduction}

The dynamics of viscous melts and solutions in the cylindrical geometry (for example, in cylindrical tubes) is of interest for different technical applications and natural phenomena. In particular, the polymeric melts are used in chemical technologies and confectionery industry, the blood flow in the vessels is often described as the Poiseuille flow of polymeric liquid \cite{lev}-\cite{gs}. As a rule, such flows are considered in incompressible approximation by neglecting the possible changes in the polymer liquid (see, for example, Refs.          \cite{lev,de,gs} and references therein). 

However, there are such physical situations where these assumptions are unacceptable.  In particular, a typical example of such systems is polymeric flow with some chemical reactions, for example, destruction and fusion of polymer chains when one should consider the flow in a compressible limit \cite{lev,uk,de}. In the present piece, we will study such an viscous flow with such chemical reactions in two dimensional cylindrical geometry. 

Common property of all these systems is the follow: The nonstationary nature of the processes occurring in such media and the self nonlinearity of the systems define the complex evolution. As a result, the basic features of these systems is their high sensitivity to external factors including perturbations in initial and boundary conditions since minor external perturbations can lead to very pronounced changes in the system dynamics (see, for example, Refs.\cite{np} - \cite{ks} and references therein). 

In order to clarify this point one should use a fully nonlinear treatment to investigate the evolution of this system. In general, however, such a statement is an unsolvable problem. So we have restricted our consideration to a particularly simple form of nonlinear dynamics of the viscous, time-dependent flow which can be studied analytically. Here we present a class of exact solutions of the fully nonlinear hydrodynamic equations describing the compressible, viscous flow which may be useful to determine the direction of the real system behavior. Based on this description, we have compared the pattern formation in the dissipative flows with the corresponding non-dissipative evolution the flow having the same origin parameters.

\section{The flow model}

We shall study the pattern formation in axially symmetric ($\partial_{\varphi}=0$), viscous flow consisting from the particles of one kind $A$ where there occur some chemical reaction which has changed the flow composition. For simplicity, we shall ignore the  friction force and the influence of chemical source on the full flow momentum. Also, our consideration will be limited to the Newtonian compressible but isentropic and spatially homogeneous liquid when the flow density is a function of time, $n_A=n_A(t)$. In this simplest case there is no influence of the pressure gradient since $\nabla p \equiv 0$. Furthermore, ignoring by internal structure of the liquid particles, we shall neglect the bulk viscosity.

So the dynamics of such fluid is governed by the Navier-Stokes equation
\begin{equation}
m_A n_A\left(\partial_t \mathbf{V}_A + (\mathbf{V}_A \cdot \nabla) \mathbf{V}_A \right)= \eta \Delta\mathbf{V}_A + {\eta\over 3}\nabla(\nabla\cdot\mathbf{V}_A)  \/, 
\label{1_ac}
\end{equation}
where $\mathbf{V}_A$ is the  velocity, $m_A$ is the mass of particles for the movable fluid; here the coefficient of dynamic viscosity $\eta$ is assumed to be  constant. The change of the flow density is defined by the equation of continuity with the source term: 
\begin{equation}
\partial_t n_A + n_A \nabla \cdot \mathbf{V}_A =W\/,
\label{2_ac}
\end{equation}
where $W$ is the rate of chemical reaction;  the specific form  of $W$ will be defined further.
 
It is convenient to rewrite the system (\ref{1_ac})-(\ref{2_ac}) 
in  dimensionless form. Normalizing the densities by the initial density $n_{A0}=n_A(t=0)$ the coordinates by an arbitrary space scale, say the initial size of initial radius of flow $b_0=b(t=0)$, velocity by the initial velocity $V_0=V_z(t=0)$, the time  by the time scale $b_0/V_0$ and the velocity of chemical reaction by the value $n_{A0} W_0$, where $W_0$ is the characteristic  collision frequency of nonelastic process, we arrive at the following system of equations:
\begin{equation}
\partial_t \mathbf{V}_A + (\mathbf{V}_A \cdot \nabla) \mathbf{V}_A = {1\over Re} \left[ \Delta\mathbf{V}_A + {1\over 3}\nabla(\nabla\cdot\mathbf{V}_A)\right]\/, 
\label{5_ac}
\end{equation}
\begin{equation}
\partial_t n_A + \nabla \cdot (n_A \mathbf{V}) =-Ch W\/,
\label{6_ac}
\end{equation}
here $Re=\eta/(b_0 m_A n_{A0} V_0)$ is the  Reynolds number which denotes the ratio of the nonlinear inertial term to the viscous dissipative term in Eq. (\ref{5_ac}), and $Ch=n_0 V_0/b_0 W_0$ is the dimensionless number which  defines the ratio between the characteristic kinetic time of the system $\tau_{ch}=n_{A0}/W_0$  and the characteristic time for the process of macroscopic transfer, in the present case it is $\tau_{tr}=b_0/v_0$.

In order to solve the system (\ref{5_ac})-(\ref{6_ac}) without perturbation, we look for self-similar flow structures in the form
\begin{equation}
\mathbf{V}_A = {A(t) \over 2} r \mathbf{e}_r +   \left[C(t)+ B(t)z + H(t)r^2\right] \mathbf{e}_z\/,
\label{9_ac}
\end{equation}
where $A(t)$, $C(t)$, $B(t)$ and $H(t)$ are associated with the radial and axial velocity components and are still to be determined, and the density dependence  corresponds to spatially uniform medium.  Clearly, the simple basis structure proposed in Eq. (\ref{9_ac}) is not unique, however, depending on the initial data $A(t=0)=A_0$, $B(t=0)=B_0$, $C(t=0)=C_0$ and $H(t=0)=H_0$, this relation can describe  complex and physically interesting behavior of the system.

Since the density does not depend on the space coordinates the expression for chemical rate $W$ has no spatial dependence too.  So substitution of Eq. (\ref{9_ac}) into Eq. (\ref{6_ac}) yields
\begin{equation}
\dot{n}_A + (A+B) n_A =-Ch W \/. 
\label{10_ac}
\end{equation}
Inserting Eq. (\ref{9_ac}) into Eq. (\ref{5_ac}) and taking into account that in our model $dp/dr=0$, we obtain
$$\left(\dot{A}+{A^2\over 2}\right)r=0$$
and 
$$\dot{C}+BC-{4\over Re}H +\left(\dot{B}+B^2\right)z + \left(\dot{H}+(A+B)H\right)r^2=0\/.$$
which are valid for any $r$ and $z$. Thus the coefficients for each power of $r$ and $z$ should be equated to zero, namely
\begin{equation}
\dot{A}+{A^2\over 2}=0\/,
\label{11_ac}
\end{equation}
\begin{equation}
\dot{B}+B^2=0\/,
\label{12_ac}
\end{equation}
\begin{equation}
\dot{H}+(A+B)H=0\/,
\label{13_ac}
\end{equation}
\begin{equation}
\dot{C}+BC={4\over Re}H\/.
\label{14_ac}
\end{equation}
Thus, the spatial and temporal parts of all physical values are kept apart. That is one of simplest form of nonlinear dynamical system whose behavior depends on the parameters $Re$ and $Ch$. So it is of interest to study the influence of each of these parameters on the dynamics of the system separately.  

\section{Dynamics for $Ch \to 0$}

First, we consider the primitive dynamical properties of the profile (\ref{9_ac}) inherent in the limit $Re \to \infty$ and $Ch \to 0$. These peculiarities will always be significant at the initial stage, when the dynamics of the flow is only determined by the cylindrical geometry.  

Although in the present  case t $V_{\varphi}=0$, in this flow field there appears an azimuthal component of the vorticity, 
\begin{equation}
\vec{\omega} =-2H(t)r{\bf e}_{\varphi}\/, 
\label{15_ac}
\end{equation}
which confirms that there is rotation in the plane $\varphi=$const. In order to demonstrate the nature of vortex origin in this flow field one should rewrite Eq. (\ref{5_ac}) as
\begin{equation}
\partial_t \vec{\omega} + (\mathbf{V}_A \cdot \nabla) \vec{\omega} - (\vec{\omega} \cdot \nabla) \mathbf{V}_A = 0
\label{omeg_ac}
\end{equation}
It is easy to see that term $(\vec{\omega}_s \cdot \nabla) \mathbf{V}_A \equiv 0$ for (\ref{9_ac}) and (\ref{15_ac}). It implies that the value of vorticity $\vec{\omega}$ is  preserved along the trajectory described by Eq. (\ref{omeg_ac}).
Now, let us proceed to considering what happens if we have a finite $Re$.
From Eqs. (\ref{10_ac})-(\ref{13_ac}) we get
\begin{equation}
n_A=n_{A0}e^{-\theta}, \hspace{7mm} H=H_0e^{-\theta}\/, 
\label{16_ac}
\end{equation}
and value
\begin{equation}
\theta(t)=\int_0^t(A(z)+B(z))dz\/, 
\label{17_ac}
\end{equation}
here
\begin{equation}
A(t)={A_0 \over 1+A_0t/2}\/, \hspace{7mm} B(t)={ B_0 \over 1+B_0t}\/.
\label{18_ac}
\end{equation}
Noting if $A_0 <0$ or $B_0 <0$, then Eq. (\ref{18_ac}) describes an approach to a singularity of flow. That is a well-known intrinsic feature of fluids especially in inviscid, pressureless limit.

Substituting  (\ref{18_ac}) into (\ref{16_ac}), we arrive at
\begin{equation}
n_A(t)={n_{A0} \over (1+B_0t)(1+A_0t/2)^2}, \hspace{7mm} H(t)={H_0 \over (1+B_0t)(1+A_0t/2)^2}\/. 
\label{19_ac}
\end{equation}
Inserting $H$ from (\ref{19_ac})  and $B$ from (\ref{18_ac}) into (\ref{14_ac}), we obtain
\begin{equation}
C(t)=\left[ {8H_0t \over Re(2+A_0t)} +C_0\right]{1 \over (1+B_0t)}\/. 
\label{20_ac}
\end{equation}
Relations (\ref{19_ac}) and (\ref{20_ac}) can be considered as the basic dynamical flow structure of the model worked out.
\begin{figure}
\centering
\includegraphics[width=4.in]{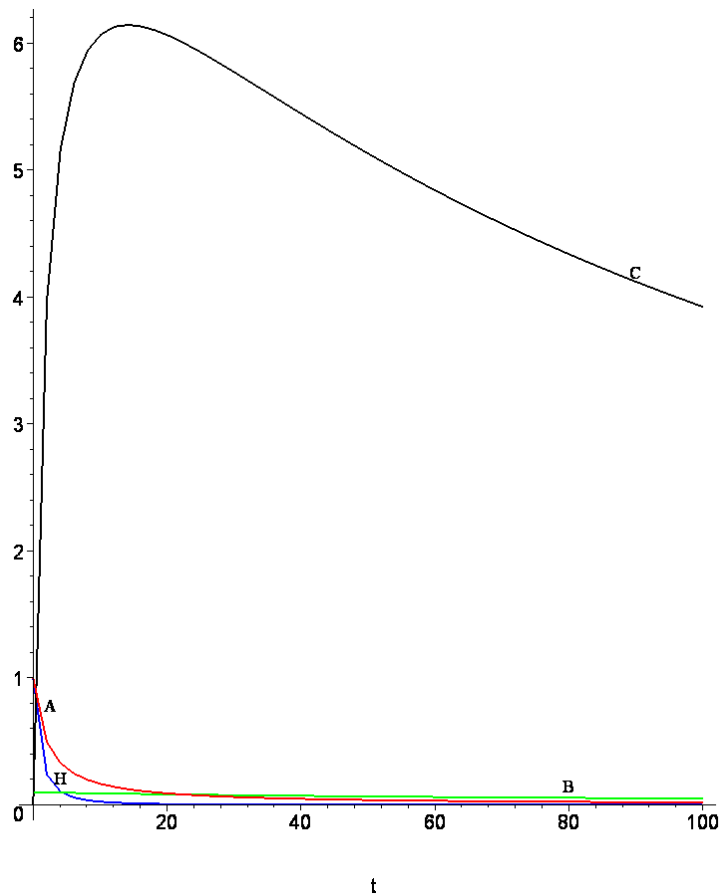}
\caption{The evolution of flow for $A_0=1$, $B_0=10^{-2}$, $C_0=10^{-3}$,
$H_0=1$ under $Re=1$}
\label{f_2}
\end{figure}

\begin{figure}
\centering
\includegraphics[width=4.in]{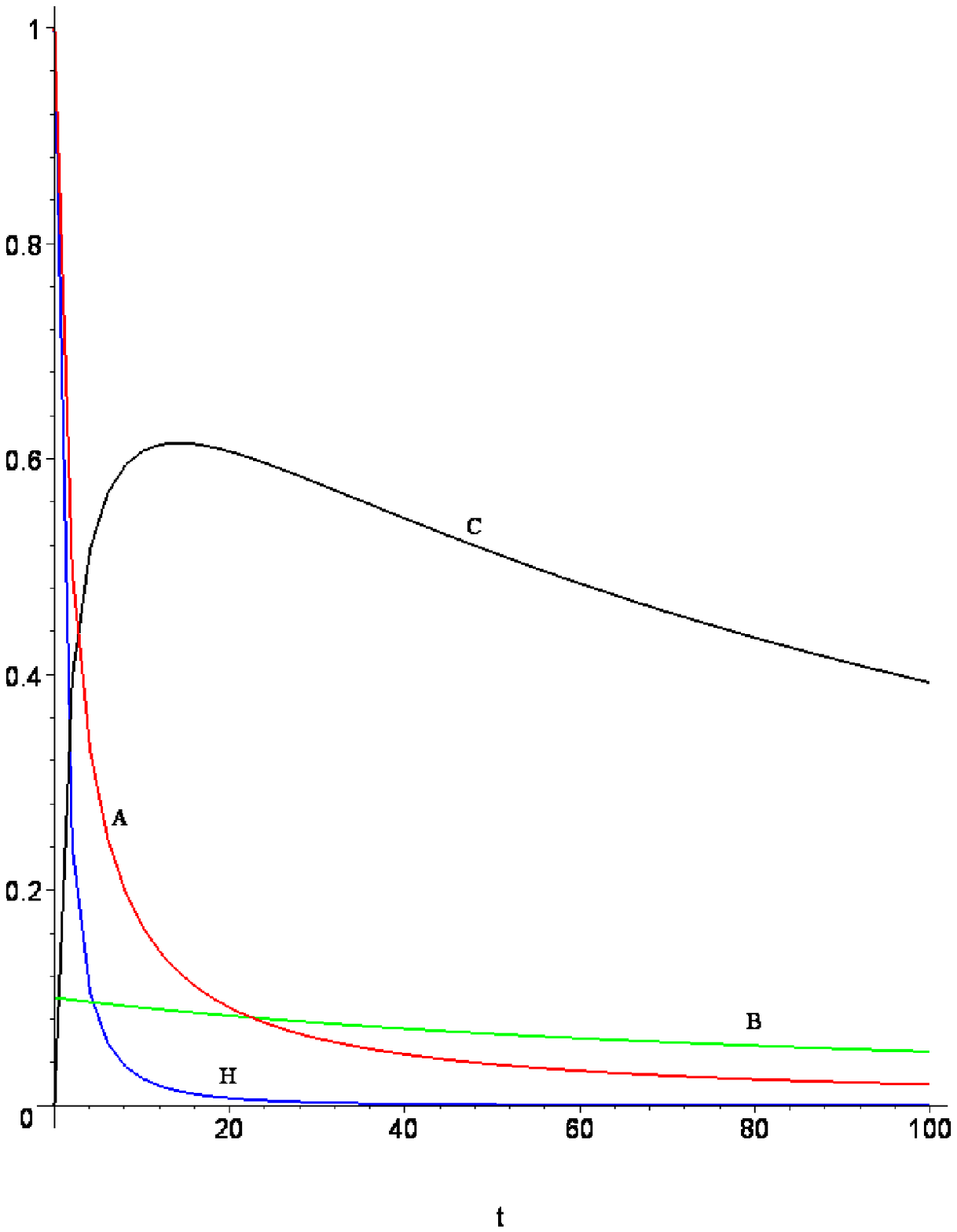}
\caption{The evolution of flow for $A_0=1$, $B_0=10^{-2}$, $C_0=10^{-3}$, $H_0=1$ under $Re=10$}
\label{f_6}
\end{figure}

It is worth touching upon the physical sense of the initial data in present case. As is seen from (\ref{18_ac})-(\ref{20_ac}), here the radial and axial flow components are strongly coupled, especially for $A_0<0$ and $B_0 <0$, when there is flow compression in the radial direction. On the other hand, for $A_0>0$ and $B_0>0$, there is no such effect since $H(t)$ decreases with time and there is no enhancement of axial flow. As is seen from (\ref{20_ac}), there is a strong influence the Reynolds number on a temporal component of axial velocity $C(t)$. Although, as is seen from the above dependences, the characteristic value of $C$ is determined by the initial values $A_0$, $B_0$ and $H_0$. It means that in practical estimations one should take into account only  component $C(t)$ by neglecting  other components. However, such statement is valid only for the expanding flows and we have to take into account the effect of initial values $A_0$, $B_0$ and $H_0$.

As an illustration of such dynamics in Fig. \ref{f_2} and \ref{f_6}  we present some numerical solutions of (\ref{14_ac}) for $Re=1$ and $Re=10$  respectively. Figure \ref{f_2} is similar to Fig. \ref{f_6}, they differ only in the magnitude of the value $C(t)$.   The present pictures show 
that there exists a strong influence of the initial data and the Reynolds number  for $Re>1$ when we should take into account  the full structure of (\ref{9_ac}).

\section{Dynamics for ($Ch\ne0$)}

We now consider the flow with chemical reactions  when $Ch\ne0$. As is seen from (\ref{10_ac})--(\ref{14_ac}), in our simplest model there is an effect of velocity field on densities but not vice versa. Such behavior comes about because of the absent of the terms corresponding nonelastic process (e.g. chemical sources, friction force) in Eq. (\ref{1_ac}) which may change the full flow momentum. In order show these features in the most simple way we restrict our consideration by a particularly simple  chemical kinetics by taking the reactions of first order only. However, it should be borne in mind that such simple conditions can be realized in some practical cases.

As a typical example of a first-order reaction, one can point out the decomposition reaction of sucrose (labeled as $A$) to fructose (labeled as  $B$) and glucose (labeled as $C$) occurring in the sucrose melt (see, for example, \cite{ut}):
\begin{equation}
A \to B + C \/.$$
\label{d_ac}
\end{equation}
In this case, putting the flow density being equal to the density of sucrose melt, we can write the rate of chemical reaction as $W=W_0 n_A$. Then Eq. (\ref{10_ac}) is reduced into
\begin{equation}
\dot{n}_A + (A+B+Ch)n_A =0 \/, 
\label{21_ac}
\end{equation}
where the functions $A(t)$ and $B(t)$ are determined by the relations (\ref{18_ac}). The solution of this equation is 
\begin{equation}
n_A(t)={n_{A0} \exp(-Ch t)\over (1+B_0t)(1+A_0t/2)^2}\/. 
\label{22_ac}
\end{equation}

As is seen from (\ref{22_ac}), the chemical reaction affects the flow via the parameters $A_0$, $B_0$ and $Ch$. In absence of a chemical reaction ($Ch \to 0$), this relation describes the flow by inertia in collisionless  media. In the case $Ch \to 0$, depending on the values $A_0$, $B_0$, it is possible to obtain either an acceleration or a slowing down of the reaction rate. That is that we observe new quality – the influence of the velocity parameters ($A_0$, $B_0$) and  kinetic number ($Ch$) on the formation of density structure.

Now we consider the case when the reaction rate is defined by the external conditions. Let the system be the medium containing movable, particles of type $A$ (e.g. that can be particles of sucrose, glucose or fructose) and immobile particles of the $B$ type (e.g. this may be an additional fruit fraction) between which there is some chemical reaction leading to the formation of new immobile component $AB$: 
\begin{equation}
A + B \to AB \/.
\label{s_ac}
\end{equation}
Such a situation can be realized when $m_A \ll m_B$, here $m_B$ is the mass of $B$ particle. In particular, this kinetic scheme can be considered as a process of complicating the structure of the polymer macromolecule to a scheme which is similar to the reaction (\ref{d_ac}) (see, for example, \cite{ut}). Also, we assume that the $A$ component is in excess so that $n_A \gg n_B$. In this case, one can neglect the change of the $A$ component and one can write $W=W_0 n_B$. Then the governing equations are 
\begin{equation}
\dot{n}_A + (A+B)n_A = -Ch n_B \/, 
\label{23_ac}
\end{equation}
\begin{equation}
\partial_t n_B =-Chn_B\/.
\label{24_ac}
\end{equation}
From these equations, we can get the relation for $n_{A}$: 
\begin{equation}
n_A(t)={n_{A0} + Chn_{B0} \int_0^t(1+B_0z)(1+A_0z/2)^2\exp(-Ch z)dz\over (1+B_0t)(1+A_0t/2)^2}\/, 
\label{25_ac}
\end{equation}
where $n_{B0}=n_B(t=0)$. 

Comparing the relations (\ref{22_ac}) and (\ref{25_ac}) with Eq. (\ref{19_ac}), one can come to the conclusion  that the dependence of $n_A$ for $Ch \ne 0$ in outline corresponds to a dynamic transformation of the basic structure (\ref{19_ac}). As is seen from these relations, in all cases when $ A_0<0$ and $B_0 <0$, our model predicts a singular behavior for finite time. That is, our pressureless model becomes invalid after time $\tau_c= {\rm min}(1/|A_0|, 1/|B_0|)$.

It would be of interest to discuss this peculiarity from the point of view accelerating chemical reactions near collapse-like points in the cylindrical geometry. In our case, the present results for reactions of first order indicate that there is a relatively weak effect of chemical terms on the flow dynamics since the nonlinearity is too strong that the system experiences a collapse-like behavior for small times. Proceeding from this point one may expect that singularities form cellular structures in cylindrical geometry. As a result, the intensity of the chemical reactions strongly can increase at these points. Such special features are expected to play an important role, in particular, for reactions of decomposition considered in the section 4.  However, it is only our assumption or guess and no more. In order to show the realization of this hypothetical mechanism we should study the above-outlined script of dynamics for dissipative flows with a chemical reaction in multidimensional geometry. 

\section{Concluding remarks}

In this paper we have studied  the dynamics of the viscous, two-dimensional flow with chemical reactions. In this case  we have to take into account the change of flow density, i.e. we have to consider the flow dynamics for compressible medium. In order to get a full analytical description of such issue we have studied only spatially homogeneous flows of the form (\ref{9_ac}) and we have considered the reaction of first order only.  These reactions belong to the simplest type of possible reactions. However, the present results indicate that the same features can be observed in more complex systems.  

We have shown that  the density collapse  may occur in the flow under analysis [see Eqs. (\ref{22_ac}) and (\ref{25_ac})] similarly to what happens in a dissipative or plasma flow [see, for example, \cite{ksch,kbk}). It is important to stress that the above-described behavior can be observed only for the initial conditions and the parameters stated in Eq. (\ref{9_ac}). However, in the present case the evolution of the flow is not restricted by any physical mechanism, such as the pressure gradient, which is usually presumed to limit the growth of the density peak. However, this problem is still under investigation.

In conclusion, we would like to emphasize that one can expect the significant acceleration of kinetic reactions at these singular points. Thus, the above results suggest  that  the rate of kinetic reactions can be controlled by changing the initial or boundary conditions in some flows of polymer liquid in various industrial installations \cite{ut,mar,ae}.\\
The work of A.R.K was supported by the Ministry of Science and Education of the Russian Federation under No. 14.575.21.0169 (RFMEFI57517X0169).

\end{document}